\documentstyle[aps,prl,floats,epsf]{revtex}

\newcommand{\bastar}{\begin{eqnarray*}}
\newcommand{\eastar}{\end{eqnarray*}}
\newskip\humongous \humongous=0pt plus 1000pt minus 1000pt

\newif\ifdtup

\relax

\draft
\preprint{SNUTP 01/013}
\begin{document}
\twocolumn[\hsize\textwidth\columnwidth\hsize\csname@twocolumnfalse%
\endcsname
\title{\bf Localized gravity and mass hierarchy in $D=6$
with the Gauss-Bonnet term}
\bigskip

\author{Jihn E. Kim,
Bumseok Kyae
and Hyun Min Lee}
\address{ Department of Physics and Center for Theoretical Physics,
Seoul National University,
Seoul 151-747, Korea \\
{\scriptsize jekim@phyp.snu.ac.kr,
kyae@fire.snu.ac.kr, minlee@phya.snu.ac.kr}
\\ \vskip 0.3cm
}
\maketitle

\begin{abstract}
We obtain the localized gravity on the intersection of two orthogonal 
{\it non-solitonic} or {\it solitonic} 4-branes 
in $D=6$ in the presence of the Gauss-Bonnet term. The tension of the 
intersection is allowed to exist unlike the case without the Gauss-Bonnet term. 
We show that gravity could be confined to the solitonic 4-branes 
for a particular choice of the Gauss-Bonnet coupling.
If the extra dimensions are
compactified with the $T^2/(Z_2\times Z_2)$ orbifold symmetry, the
mass hierarchy between the Planck scale and the weak
scale can be explained by putting our universe at the TeV intersection of 
positive tension located at the orbifold fixed point.

\vspace{0.3cm}
\end{abstract}

\narrowtext
\bigskip
                        ]
\def\L{$\Lambda$}
\def\F{$\{F\tilde F\}$}
\def\p{\partial}

\section{introduction}

The recent proposals for the fundamental TeV scale physics\cite{ADD,RS1}
have been a great surprise in high energy physics, which has not been 
noted for a long period of superstring research.
Of particular interest is the first Randall-Sundrum(RSI) model\cite{RS1}
in which the warp factor geometry in the extra direction in the 5
dimensional spacetime(5D) introduces a large exponential suppression factor, 
enabling one to introduce a TeV scale from the Planck scale with 
an O(10) ratio of the input parameters, through the compactification
of the extra dimension $y$ on $S_1/Z_2$. There are two branes in the
RSI model: the brane 1 (B1) located at $y=0$ and brane 2 (B2) located at 
$y=y_2$.
  
Probably a more interesting proposal is the second Randall-Sundrum(RSII)
model\cite{RS2} in which only one brane (B1) located at $y=0$ is introduced.
Thus, the fifth dimension is not compactified, but still this model
can describe a meaningful effective 4 dimensional(4D) physics since
the gravity is localized around B1. It is an alternative to
compactification idea of the extra dimension(s). Both
Randall-Sundrum models need AdS spacetime in the bulk.
 
Subsequently, extensions of the RS type models were proposed 
toward the hierarchy solution\cite{cohen,chacko,KKL1,moon},
for the study of localization of gravity
\cite{ADDK,KL,gregory,vil,tony,moon}, and for other 
aspects\cite{rsref,kaku1,kaku2}.
In particular, the RSII model have been studied toward finding
a self-tuning solution of the cosmological constant problem. 
It is because, from 
the beginning of these proposals, the solution
of the cosmological constant was sought for in the RS models since the
Einstein equations can choose a flat space even with a negative
nonvanishing bulk cosmological constant and nonvanishing
brane tension(s). But in the first proposals, the nonvanishing 
parameters should be fine tuned for the universe to be flat in 
the model\cite{RS1,RS2}. It has been suggested 
that introduction of a bulk real scalar field with a coupling to the brane 
may give a self-tuning of the cosmological constant but it retains 
the serious fine tuning problem due to a naked 
singularity\cite{self}. There exists an example for the selftuning solution
with an unconventional interaction of a bulk antisymmetric tensor field 
\cite{KKL2} which may shed more light toward a final solution of
the cosmological constant problem. We note that the final 
solution must allow inflation which
seems to be needed for the explanation of homegeneity and isotropy of
the observed universe\cite{guth}.

In the intersecting brane world scenarios in higher than five dimensions
\cite{ADDK,branes}, our 
universe is regarded as a 3-brane with higher codimensions given by the 
common intersection of higher dimensional objects with lower codimensions. 
However, when we consider
discrete sources of higher dimensional objects in the bulk space, 
no additional contribution is allowed from the brane-brane interaction to 
the tension of their intersection corresponding to the 3-brane tension, 
since the Einstein tensor just gives rise to one-dimensional delta function 
from the intersecting branes. This behavior is well understood in the smooth 
limit of intersecting branes. For instance, for $n$ orthogonal $(n+2)$-branes 
in $D=4+n$, each $(n+2)$-brane has the tension $T_{n+2}=M^{n+4}L$ while the 
tension of their intersection is $T_3=M^{n+4}L^n$ by dimensional analysis. 
($M$ is the $(4+n)$-dimensional fundamental scale and $L$ is the brane 
thinkness.)
Therefore, the 3-brane tension shows up with higher power of $L$, so it gets 
suppressed in the thin brane limit, which means that higher
curvature terms should be taken into account for better resolution to see such a
thin 3-brane.  
Without nonzero 3-brane tension, it is difficult to discuss on the generation 
of vacuum energy after phase transition on the intersection as our world. 
Because the corresponding 
nonzero tension of the intersection is not allowed, the vacuum 
energy induced by phase transition has no way but at most to  
leak away along the intersecting branes, whose tensions are allowed to be 
nonzero. In this context, it is necessary that the nonzero brane-brane 
interaction or the nonzero 
tension of the intersection should appear in a natural way.

In this paper we consider the RS type solution for the case of two orthogonally
intersecting {\it non-solitonic} 4-branes\footnote {Here, {\it solitonic} means 
being 
supported by gravity only while {\it non-solitonic} does by sources.} and one 3-brane (or string) 
on their intersection 
in D=6 when the Gauss-Bonnet term is added in the bulk action. 
In that case, we can regard our world
as a common intersection of two 4-branes where the localization of gravity 
arises. In the existence of the Gauss-Bonnet term, in particular, 
a string tension should
be introduced at the beginning to match an additional boundary condition on 
the intersection. So, our solution with two 4-branes and one string is based 
on two fine-tuning conditions between input parameters but 
there is a possibility for naturally regarding 
the vacuum energy in our world 
as the string tension in the intersecting brane world scenario, 
which has not been possible to get without the Gauss-Bonnet term. 
Thus, it seems 
that the higher curvature terms know about the inner structure of the 
intersecting branes whilest the Einstein-Hilbert term has the lower resolution.  

For a special relation between the bulk cosmological constant and the 
Gauss-Bonnet coupling in our model, it is shown that
there exists a string solution with codimension-2 
by considering the $Z_2\times Z_2$ symmetry of the extra dimensions 
as usually imposed in the case of the two orthogonal 4-branes. 
In that case, the bulk space is found to be a discrete patch  
of the pure $AdS_6$ space to make the bulk symmetry manifest 
and the resultant discontinuities of the derivative of the metric across 
the symmetry axes are shown to be automatically cancelled between those 
derived from 
the Einstein-Hilbert term and the Gauss-Bonnet term in the equations of motion 
without the need of introducing 4-branes along the symmetry axes. 
In other 
words, it is shown that the Einstein-Gauss-Bonnet gravity itself 
is able to support singularities produced on orbifolding without the need of 
introducing additional {\it non-solitonic} singular sources. From 
the point of view of the Einstein's gravity, however, the 
singularities are interpreted as the so called {\it solitonic} 
4-branes\cite{kaku2}, 
of which tensions are determined 
by the Gauss-Bonnet coupling and the 3-brane tension. 
Nonetheless, since the solitonic 4-branes are supported by gravity only without 
sources in $D=6$, they don't give any fine-tuning conditions.
Therefore, on patching the $AdS_6$ bulk in the $Z_2\times Z_2$ invariant way, 
there exists a solution of a string residing on the intersection 
of two 
solitonic 4-branes, which is based on one fine-tuning condition between bulk 
parameters but for which the 3-brane cosmological constant
$\Lambda_1$ can take any positive value 
without being involved in any fine-tuning. In particular, it is interesting to 
see that there arises the confinement of gravity to the solitonic 4-branes, 
which results in exactly two copies of the 5D RSII model in $D=6$.  
 
For the string solution with codimension 2 in $D=6$, it has been shown that the singular global string solution is possible
with a massless scalar field in the flat bulk by
the unitarity boundary condition at the singularity\cite{cohen}.
Later, it was pointed
out that there exist regular global string solutions 
by introducing a bulk cosmological constant\cite{gregory,vil,moon}. 
One more interesting observation is that the local string defects 
were shown to have the localized gravity with no fine-tuning of the 
bulk cosmological constant, but here
the components of the string tension are required to satisfy 
a certain relation\cite{tony}, which is a fine tuning. 
Pertinent to our study of this paper,
we note the work of Corradini and Kakushadze
in which it has been argued that it is 
possible to have the localized gravity on a solitonic 3-brane with the 
Gauss-Bonnet term while freely choosing the brane cosmological constant 
equivalent to a deficit angle in the extra polar coordinate
in the 5th and the 6th space\cite{CK}. (Note that a similar
result was known in the case with 3-brane sources in the 6D Einstein gravity 
without a bulk cosmological constant\cite{sundrum}
and with a positive bulk cosmological constant\cite{chodos}.) 
This solution has 
one fine-tuning condition between bulk parameters and there exists 
a conical singularity corresponding to the brane tension\cite{CK}.

Based on our string solution in the intersecting brane scenario, we can 
compactify the extra dimensions with 
the $T^2/(Z_2\times Z_2)$ orbifold symmetry. 
Then, we can show that the hierarchy problem can be solved if we put
the branes at the four fixed points of the orbifold $T^2/(Z_2\times Z_2)$
and the neighboring two 3-branes are connected to each other by one 4-brane. 
In this case, the positive tension brane diagonally 
far away from the origin of the extra dimension is regarded as our 
universe and some other three 3-branes as the hidden branes. 

In Sec. II, we obtain a 3-brane (or string) solution in the
EGB theory. It is the most relevant generalization of the
RSII model. There appears {\it solitonic} 4-brane solutions. 
In Sec. III, we consider the metric perturbation near the background geometry
and ensure that there is no tachyonic mode of graviton. Then, in Sec. IV, we 
make discussions on the gravity confinement to the solitonic 4-branes.
In Sec. V, we
compactify 6D with the $T^2/(Z_2\times Z_2)$ orbifold symmetry
and obtain four fixed points where 3-brane sources can be
placed. It is the most relevant generalization of the
RSI model in which a TeV 3-brane can occur naturally.
Sec. VI is a conclusion. 

\section{localized gravity on a 3-brane in 6D}   

If we impose the $Z_2$ symmetry on each extra dimension in $D=n+4$ dimensional
generalization of the RS model, we should have $(n+2)$-branes orthogonally 
intersecting to each other to match the boundary conditions of the metric
\cite{ADDK,branes}. Therefore, the 3-brane as our universe only appears as
the common intersection of all the $(n+2)$-branes\cite{ADDK}, but without its 
tension. However, 
in the presence of the Gauss-Bonnet term, from which no higher than 
second derivatives are derived in the equations of motion, 
the intersection of two orthogonal 4-branes in $D=6$ is required to have a 
nonzero tension, which will be shown below. 

When the Gauss-Bonnet term is added as the next leading-order 
ghost-free interaction to the Einstein-Hilbert term 
in 6 dimension(6D) with two spacelike extra dimensions, we start with the 
Einstein-Gauss-Bonnet(EGB) 6D action with singular brane sources,
\begin{eqnarray}
S_6&=&\int d^4x dz_1 dz_2\sqrt{-g}\bigg[ {M^4\over 2}R-\Lambda_b \nonumber \\ 
&+&\frac{1}{2}\alpha M^2 (R^2
-4R_{MN}R^{MN}+R_{MNPQ}R^{MNPQ})\bigg] \nonumber \\
&+&\int d^4xdz_2 \sqrt{-g^{(z_1=0)}}(-\Lambda_{z_1})\nonumber\\
&+&\int d^4xdz_1 \sqrt{-g^{(z_2=0)}}(-\Lambda_{z_2})\nonumber \\
&+&\int d^4x\sqrt{-g^{(z_1=0,z_2=0)}}(-\Lambda_1)
\label{action} 
\end{eqnarray}
where $g, g^{(z_1=0)},g^{(z_2=0)}$ and $g^{(z_1=0,z_2=0)}$ 
are the determinants of the metrics in the
bulk, orthogonally intersecting 4-branes and a 3-brane, 
$M$ is the six dimensional gravitational constant, $\Lambda_b$,
$\Lambda_{z_1},\Lambda_{z_2}$, and $\Lambda_1$ are the bulk 
and the brane cosmological constants,
$\alpha$ is the effective coupling. We considered the 4-branes to
write down general equations of motion, but we will see later that
there is a possibility of getting the string solution without these 4-brane 
sources by imposing the $Z_2\times Z_2$ symmetry in the bulk.

Equations of motion in this EGB theory are,
\begin{eqnarray}
G_{MN}+H_{MN}=M^{-4}T_{MN}.
\end{eqnarray}
The tensors in the above equation are 
\begin{eqnarray}
G_{MN}&\equiv&R_{MN}-\frac{1}{2}g_{MN}R,\label{einstein} \\
H_{MN}&\equiv&\frac{\alpha}{M^2}\bigg[-\frac{1}{2}g_{MN}(R^2-4R^2_{PQ}
+R_{PQST}R^{PQST}) \nonumber \\
&+&2RR_{MN}-4R_{MP}R_N\,^P-4R^K\,_{MPN}R_K\,^P \nonumber \\
&+&2R_{MQSP}R_N\,^{QSP}\bigg], 
\label{hcurv}\\
T_{MN}&\equiv&-\Lambda_b g_{MN}
-\frac{\sqrt{-g^{(z_1=0)}}}{\sqrt{-g}}\Lambda_{z_1} \delta(z_1)
\delta^p_M\delta^q_N g^{(z_1=0)}_{pq}\nonumber \\ 
&-&\frac{\sqrt{-g^{(z_2=0)}}}{\sqrt{-g}}\Lambda_{z_2} \delta(z_2)
\delta^a_M\delta^b_N g^{(z_2=0)}_{ab}\nonumber \\
&-&\frac{\sqrt{-g^{(z_1=0,z_2=0)}}}{\sqrt{-g}}\Lambda_1\delta(z_1) \delta(z_2)
\delta^\mu_M\delta^\nu_N g^{(z_1=0,z_2=0)}_{\mu\nu},
\end{eqnarray}
where the indices $M,N=(0,1,2,3,5,6)$, $p,q=(0,1,2,3,6)$, 
$a,b=(0,1,2,3,5)$ and $\mu,\nu=(0,1,2,3)$.

Taking the metric ansatz as a conformally flat one in 6D,
which is manifestly 4D Poincar$\grave{e}$ invariant,
\begin{eqnarray}
ds^2_6&=&A^2(z_1,z_2)(\eta_{\mu\nu}dx^\mu dx^\nu+dz_1^2+dz_2^2),\label{metric}
\end{eqnarray}
where $(\eta_{\mu\nu})=diag.(-1,+1,+1,+1)$,
we obtain the tensor components $G_{MN}$ and $H_{MN}$ as follows,
\begin{eqnarray}
G_\mu\,^\nu&=&\frac{2}{A^2}\bigg[\bigg(\frac{A'}{A}\bigg)^2
+\bigg(\frac{\dot{A}}{A}\bigg)^2+2\frac{A^{\prime\prime}}{A}
+2\frac{\ddot{A}}{A}\bigg]\delta_\mu^\nu,\\
G_{5}\,^{5}&=&\frac{2}{A^2}\bigg[5\bigg(\frac{A'}{A}\bigg)^2
+\bigg(\frac{\dot{A}}{A}\bigg)^2+2\frac{\ddot{A}}{A}\bigg],\\
G_{5}\,^{6}&=&\frac{4}{A^2}\bigg[-\frac{\dot{A}'}{A}
+2\frac{\dot{A}A'}{A^2}\bigg],\\
G_{6}\,^{6}&=&\frac{2}{A^2}\bigg[5\bigg(\frac{\dot{A}}{A}\bigg)^2
+\bigg(\frac{A'}{A}\bigg)^2+2\frac{A^{\prime\prime}}{A}\bigg],
\end{eqnarray}
and 
\begin{eqnarray}
H_\mu\,^\nu&=&-\frac{12\alpha}{M^2}\frac{1}{A^4}
\bigg[-3\bigg(\bigg(\frac{A'}{A}\bigg)^2
+\bigg(\frac{\dot{A}}{A}\bigg)^2\bigg)^2\nonumber\\
&+&4\bigg(\frac{A'}{A}\bigg)^2\frac{A^{\prime\prime}}{A}
+4\bigg(\frac{\dot{A}}{A}\bigg)^2\frac{\ddot{A}}{A}
+2\frac{A^{\prime\prime}\ddot{A}}{A^2}\nonumber \\
&-&2\frac{\dot{A}'}{A}\bigg(\frac{\dot{A}'}{A}
-4\frac{\dot{A}A'}{A^2}\bigg)\bigg]\delta_\mu^\nu, \\
H_{5}\,^{5}&=&\frac{12\alpha}{M^2}\frac{1}{A^4}
\bigg[-2\bigg(\frac{A'}{A}\bigg)^2\bigg(\frac{\dot{A}}{A}\bigg)^2
-5\bigg(\frac{A'}{A}\bigg)^4\nonumber \\
&+&3\bigg(\frac{\dot{A}}{A}\bigg)^4
-4\bigg(\bigg(\frac{A'}{A}\bigg)^2
+\bigg(\frac{\dot{A}}{A}\bigg)^2\bigg)\frac{\ddot{A}}{A}\bigg],\\
H_{5}\,^{6}&=&-\frac{48\alpha}{M^2}\frac{1}{A^4}
\bigg(\bigg(\frac{A'}{A}\bigg)^2
+\bigg(\frac{\dot{A}}{A}\bigg)^2\bigg)\nonumber\\
&\cdot&\bigg(-\frac{\dot{A}'}{A} +2\frac{\dot{A}A'}{A^2}\bigg), \\
H_{6}\,^{6}&=&\frac{12\alpha}{M^2}\frac{1}{A^4}
\bigg[-2\bigg(\frac{A'}{A}\bigg)^2\bigg(\frac{\dot{A}}{A}\bigg)^2
-5\bigg(\frac{\dot{A}}{A}\bigg)^4\nonumber\\
&+&3\bigg(\frac{A'}{A}\bigg)^4
-4\bigg(\bigg(\frac{A'}{A}\bigg)^2
+\bigg(\frac{\dot{A}}{A}\bigg)^2\bigg)\frac{A^{\prime\prime}}{A}\bigg]
\end{eqnarray}
where the prime and the dot denote the derivatives with respect 
to $z_1$ and $z_2$, respectively. The
energy momentum tensor $T_{MN}$ is given by
\begin{eqnarray}
T_M\,^N&=&-\Lambda_b\delta_M^N
-\frac{1}{A}\Lambda_{z_1}\delta(z_1)
\delta^p_M\delta^N_q\delta_p^q\nonumber\\
&-&\frac{1}{A}\Lambda_{z_2}\delta(z_2)\delta^a_M\delta^N_b\delta_a^b \nonumber\\
&-&\frac{1}{A^2}\Lambda_1\delta(z_1)\delta(z_2)\delta^\mu_M\delta^N_\nu
\delta_\mu^\nu.
\end{eqnarray}
Then, the (56) component of the modified Einstein's equations 
is
$$
\frac{4}{A^2}\bigg[1-\frac{12\alpha}{M^2}\frac{1}{A^2}
\bigg(\bigg(\frac{A'}{A}\bigg)^2
+\bigg(\frac{\dot{A}}{A}\bigg)^2\bigg)\bigg]
$$
\begin{equation}
\cdot\bigg(-\frac{\dot{A}'}{A}
+2\frac{\dot{A}A'}{A^2}\bigg)=0.\label{e56}
\end{equation}
Therefore, to assure that the above equation is satisfied, we require 
that the second factor vanishes,
\begin{eqnarray}
-\frac{\dot{A}'}{A}+2\frac{\dot{A}A'}{A^2}=0,\label{cond56}
\end{eqnarray}
i.e., the general solution of the metric is given by
\begin{equation}
A(z_1,z_2)\propto\frac{1}{(F(z_1)+G(z_2))}\label{solution}
\end{equation}
where $F$ and $G$ are undetermined functions of $z_1$ and $z_2$, respectively.  
Note that in case of the vanishing first factor 
in Eq.~(\ref{e56}), Eq.~(\ref{cond56}) is automatically satisfied.
To determine the exact solution of the above type, 
we can rewrite the (00)( or ($ii$)), (55) and (66) components
under the condition Eq.~(\ref{cond56}), respectively:
\begin{eqnarray}
E+e_1+e_2+e_3&=&M^{-4}\bigg[-\Lambda_b
-\frac{1}{A}\Lambda_{z_1}\delta(z_1)\nonumber \\
&&-\frac{1}{A}\Lambda_{z_2}\delta(z_2)
-\frac{1}{A^2}\Lambda_1\delta(z_1)\delta(z_2)\bigg],\label{zero}
\end{eqnarray}
\begin{equation}
E+e_2=M^{-4}\bigg[-\Lambda_b
-\frac{1}{A}\Lambda_{z_2}\delta(z_2)\bigg],\label{five}
\end{equation}
\begin{equation}
E+e_1=M^{-4}\bigg[-\Lambda_b
-\frac{1}{A}\Lambda_{z_1}\delta(z_1)\bigg],\label{six}
\end{equation}
where
$$
E=10\bigg[1-\frac{6\alpha}{M^2}\frac{1}{A^2}
\bigg(\bigg(\frac{A'}{A}\bigg)^2
+\bigg(\frac{\dot{A}}{A}\bigg)^2\bigg)\bigg]
$$
\begin{equation}
\cdot\frac{1}{A^2}\bigg[\bigg(\frac{A'}{A}\bigg)^2
+\bigg(\frac{\dot{A}}{A}\bigg)^2\bigg],
\end{equation}
\begin{equation}
e_1=\frac{4}{A}\bigg(\frac{A'}{A^2}\bigg)'\bigg[1-\frac{12\alpha}{M^2}
\frac{1}{A^2}\bigg(\bigg(\frac{A'}{A}\bigg)^2
+\bigg(\frac{\dot{A}}{A}\bigg)^2\bigg)\bigg],
\end{equation}
\begin{equation}
e_2=\frac{4}{A}\bigg(\frac{\dot{A}}{A^2}\bigg)^{\dot{\ }}
\bigg[1-\frac{12\alpha}{M^2}
\frac{1}{A^2}\bigg(\bigg(\frac{A'}{A}\bigg)^2
+\bigg(\frac{\dot{A}}{A}\bigg)^2\bigg)\bigg],
\end{equation}
\begin{equation}
e_3=-\frac{24\alpha}{M^2}\frac{1}{A^2}\bigg(\frac{A'}{A^2}\bigg)'
\bigg(\frac{\dot{A}}{A^2}\bigg)^{\dot{\ }}.
\end{equation}
Thus, the bulk equation in all the above components, $E=-\Lambda_b/M^4$, 
can be solved only if $F(z_1)=k_1z_1+c_1$ and $G(z_2)=k_2z_2+c_2$ 
($c_1,c_2$ are integration constants.), i.e.
\begin{eqnarray}
A(z_1,z_2)=\frac{1}{(k_1|z_1|+k_2|z_2|+1)}\label{sol}
\end{eqnarray} 
where the $Z_2$ symmetry is used along each extra dimension and the 
integration constants are arbitrarily chosen 
for $A$ to be 1 at $(z_1,z_2)=(0,0)$. 
$k_1,k_2$ are determined by the following relations,
\begin{equation}
k_1^2+k_2^2
=\frac{M^2}{12\alpha}\bigg[1\pm
\sqrt{1+\frac{12\alpha\Lambda_b}{5M^6}}\bigg] 
\equiv k_\pm^2,\label{k1}
\end{equation}
\begin{equation}
k_1\bigg(1-\frac{12\alpha k_\pm^2}{M^2}\bigg)
=\frac{\Lambda_{z_1}}{8M^4}, \label{k2}
\end{equation}
\begin{equation}
k_2\bigg(1-\frac{12\alpha k_\pm^2}{M^2}\bigg)
=\frac{\Lambda_{z_2}}{8M^4}, \label{k3}
\end{equation}
\begin{equation}
\alpha k_1 k_2=\frac{\Lambda_1}{96M^2},\label{k4}
\end{equation} 
where the last three equations are derived from the boundary conditions 
on the branes in Eqs.~(\ref{zero}-\ref{six}).
The first and fourth equations determine $k_1$ and $k_2$ 
in terms of $\alpha$, $\Lambda_b$, $\Lambda_1$, and 
it should be such that 
$|\Lambda_1|\leq 48|\alpha| k^2_\pm M^2$,
where the equality implies the existence of 
exchange symmetry between two extra dimensions, 
and $sign(\Lambda_1)=sign(\alpha)$ to give real solutions 
for $k_1$ and $k_2$. Then, the second and third equations 
give rise to two fine-tuning conditions between input parameters.  
Note that the Gauss-Bonnet term requires an additional 
condition, Eq.~({\ref{k4}), on the 3-brane other than those 
the Einstein-Hilbert 
action imposes on the 4-branes, Eqs.~(\ref{k2}) and (\ref{k3}). 

However, if we chose a relation between bulk parameters from the beginning,
\begin{eqnarray}  
\frac{12\alpha\Lambda_b}{5M^6}=-1\label{bulktune}
\end{eqnarray}
such that $k_\pm^2=\frac{M^2}{12\alpha}$ for $\alpha >0$, {\it non-solitonic} 
4-brane tensions 
would not be allowed to exist, viz. Eqs. (\ref{k2}) and (\ref{k3}).
Then, the 3-brane tension $\Lambda_1$ can
take any positive values without being involved in any fine-tuning 
relations. In this case, the remaining equations (\ref{k1}) and (\ref{k4}) 
just determine $k_1$ and $k_2$ in terms of $\alpha$ and $\Lambda_1$. 
This particular point in the solution space is made possible 
only with the addition of the Gauss-Bonnet term, 
but is not possible with the Einstein-Hilbert term alone.\footnote{ 
In the extension of the RS model with one extra timelike dimension 
in D=6\cite{bere}, it is shown that there exists a 3-brane 
solution as a common intersection of two 4-branes with no 
fine-tuning of the cosmological constant if the exchanging symmetry 
$y'\leftrightarrow t'$
is assumed between the extra space and time coordinates. However, 
in the existence of the Gauss-Bonnet term, there arises
a fine-tuning from the necessity of a 3-brane to match the 
boundary condition.} 
In other words, on patching the bulk space in 
the $Z_2\times Z_2$ symmetric way as shown in the chosen metric, 
we naturally obtain a string solution via 
the cancellation between those derived from the Einstein-Hilbert term and the 
Gauss-Bonnet term in the equations of motion. 
However, from the point of view of the Einstein's gravity, 
singularities on orbifolding should be seen to stem from {\it solitonic} 
4-brane tensions, just as in Iglesias and Kakushadze's\cite{kaku2}. 
In our case, the solitonic 4-brane 
tensions $f_1$($f_2$) located at $z_1=0$($z_2=0$) are determined to be positive 
as
\begin{eqnarray}
f_1=8k_1 M^4, \ \ \ f_2=8k_2 M^4 
\end{eqnarray}
where $k_1$ and $k_2$ are given by solving Eqs.~(\ref{k1}) and (\ref{k4})
under the condition Eq.~(\ref{bulktune}).
  
Then, after integrating the extra dimensions with the 4D part of the metric 
as $\bar{g}_{\mu\nu}(x)=\eta_{\mu\nu}$ in Eq.~(\ref{metric}), we obtain the 
4D effective action as follows,
\begin{eqnarray}
S_{eff}&=&\frac{M^2_{P,eff}}{2}\int d^4 x\sqrt{-\bar{g}^{(4)}}
\bigg[\bar{R}
+\frac{\alpha_{eff}}{M^2_{P,eff}}
(\bar{R}^2\nonumber \\
&-&4\bar{R}^2_{\mu\nu}
+\bar{R}^2_{\mu\nu\rho\sigma})\bigg]\label{effaction}
\end{eqnarray}
where the 4D Planck mass and the 4D Gauss-Bonnet coupling are given by
\begin{eqnarray}
M^2_{P,eff}&=&M^4\int^\infty_{-\infty} dz_1\int^\infty_{-\infty} dz_2
\bigg[A^4\bigg(1+\frac{12\alpha}{M^2}\frac{1}{A^2}
\bigg(\bigg(\frac{A'}{A}\bigg)^2\nonumber \\
&+&\bigg(\frac{\dot{A}}{A}\bigg)^2\bigg)\bigg) 
-\frac{12\alpha}{M^2}\bigg((AA')'+(A\dot{A})^{\dot{\ }}\bigg)\bigg]
\nonumber\\
&=&\frac{2M^4}{3k_1 k_2}\bigg(1+\frac{12\alpha k^2_\pm}{M^2}\bigg) 
=\frac{64\alpha M^6}{\Lambda_1}\bigg(1+\frac{12\alpha k^2_\pm}{M^2}\bigg)
\nonumber\\
&\geq& \frac{4M^4}{3k^2_\pm}\bigg(1+\frac{12\alpha k^2_\pm}{M^2}\bigg),
\label{pmass} 
\end{eqnarray}
\begin{equation} 
\alpha_{eff}=\alpha M^2\int^\infty_{-\infty} 
dz_1\int^\infty_{-\infty} dz_2 A^2 \label{alpha1}
\end{equation}  
where $(AA')'$ and $(A\dot{A})^{\dot{\ }}$ terms in the first line vanish 
after integration. 
For a negative Gauss-Bonnet coupling $\alpha$, 
the 4D Planck mass would not be positive definite 
due to the contribution from the Gauss-Bonnet term. 
Therefore, the positivity condition gives $|\alpha|< \frac{M^2}
{12k^2_\pm}$ for $\alpha<0$ and any value for $\alpha>0$.  
On the other hand, the 4D Gauss-Bonnet coupling is shown to
become logarithmically divergent after integration. This seems to be a generic 
feature of higher curvature terms, which is rephrased as the delocalization 
of gravity in warped geometry\cite{kaku1}. Nonetheless, there does not arise 
a problem in our case since 
the Gauss-Bonnet term is a total derivative in $D=4$ and thus it does not 
modify the equation of motion for graviton in the 4D spacetime. 
Therefore, we can 
drop the 4D Gauss-Bonnet term in Eq.~(\ref{effaction}) to get the 4D 
effective Einstein gravity\cite{kaku1}.

\section{metric perturbation near the background geometry}

Now that we have obtained the background solution, it is of interest to 
examine the perturbation effects of gravity near the background solution.  
Since the effects inform us how the gravitational interaction between matter
is described at low energy scales under a background geometry,   
it is indispensable to study the perturbative expansion  
and compare it with the well-known gravitational interaction.  
The perturbation in higher dimensional space-time is usually interpreted as 
the graviton in the corresponding space-time dimension, and is, 
in 6 dimension case, decomposed into 
a 4 dimensional graviton, two kinds of vectors and three kinds of scalars.  
In this section, however, we assume that the vector and scalar modes are 
decoupled by some physics due to their absence at the  
low energy scale, and we focus on the gravitational interaction 
mediated by the 4 dimensional graviton.   

Thus, for the study, let us assume the metric as the following, 
\begin{eqnarray}\label{pertmetr}
ds^2&=&\bigg[A^2(z_1,z_2)\eta_{\mu\nu}+h_{\mu\nu}(x,z_1,z_2)\bigg]
dx^{\mu}dx^{\nu} \nonumber \\
&&+A^2(z_1,z_2)(dz_1^2+dz_2^2)  \\
&=&A^2(z_1,z_2)\bigg[\bigg(\eta_{\mu\nu}+\tilde{h}_{\mu\nu}(x,z_1,z_2)\bigg)
dx^{\mu}dx^{\nu} \nonumber \\
&&+dz_1^2+dz_2^2\bigg] \label{pcmetric}~,
\end{eqnarray}
where $x$ denotes the 4 dimensional coordinate,  
and we would keep
the linear parts in $h_{\mu\nu}$ in the full expression of 
the Einstein equation.  
Here, $A(z_1,z_2)$ is the background solution given by Eq.~(\ref{sol}) and 
$h_{\mu\nu}$ represents a small perturbation near it.   
With Eq.~(\ref{pertmetr}), the linearized variations 
for $G_{\mu\nu}$, $H_{\mu\nu}$ and $T_{\mu\nu}$ are given by  
\begin{eqnarray}
\delta G_{\mu \nu}&=&-\frac{1}{2} \bigg[
\frac{1}{A^2}\Box_4
+\frac{1}{A^2}\bigg(\partial_{z_1}^{2}+\partial_{z_2}^{2}\bigg)
-26\bigg(k_1^2+k_2^2\bigg) \nonumber \\
&&+\frac{20}{A}\bigg(k_1\delta (z_1)+k_2\delta (z_2)\bigg)
\bigg]h_{\mu \nu}~, 
\\
\delta H_{\mu \nu}&=&\frac{\alpha}{M^2}\bigg[
\frac{1}{A^2}
\bigg(6(k_1^2+k_2^2)-\frac{8k_1}{A}\delta(z_1)-\frac{8k_2}{A}\delta(z_2)\bigg)
\Box_4 \nonumber \\
&&+\frac{1}{A^2}\bigg(6(k_1^2+k_2^2)-\frac{8k_2}{A}\delta(z_2)
\bigg)\partial_{z_1}^{2} \nonumber \\ 
&&+\frac{1}{A^2}\bigg(6(k_1^2+k_2^2)-\frac{8k_1}{A}\delta(z_1)
\bigg)\partial_{z_2}^{2} \nonumber \\
&&+\frac{8k_1}{A}\bigg(\frac{3k_1}{A}\delta(z_1)-\frac{k_2}{A}\delta(z_2)
\bigg)sgn(z_1)\partial_{z_1} \nonumber \\
&&+\frac{8k_2}{A}\bigg(\frac{3k_2}{A}\delta(z_2)-\frac{k_1}{A}\delta(z_1)
\bigg)sgn(z_2)\partial_{z_2} \nonumber \\
&&-96\bigg(k_1^2+k_2^2\bigg)^2 \nonumber \\
&&+\frac{k_1}{A}\delta(z_1)\bigg(168 k_1^2+152k_2^2\bigg) \nonumber \\
&&+\frac{k_2}{A}\delta(z_2)\bigg(168 k_2^2+152k_1^2\bigg) \nonumber \\
&&-160\frac{k_1k_2}{A^2}\delta(z_1)\delta(z_2)
\bigg]h_{\mu \nu}~, 
\\
\delta T_{\mu \nu}&=&-\Lambda_b h_{\mu \nu}
-\frac{1}{A}\Lambda_{z_1}\delta(z_1)h_{\mu \nu}
-\frac{1}{A}\Lambda_{z_2}\delta(z_2)h_{\mu \nu} \nonumber \\
&&-\frac{1}{A^2}\Lambda_1\delta(z_1)\delta(z_2)h_{\mu \nu}~,
\end{eqnarray}
where $\Box_4\equiv \eta^{\mu\nu}\partial_{\mu}\partial_{\nu}$, and 
we choose the traceless transverse gauge conditions, 
$\partial^{\mu}h_{\mu\nu}=h^{\mu}\,_{\mu}=0$. 

The above expressions lead to the linearized Einstein equation,  
\begin{eqnarray}
&-&\frac{1}{2A^2}\bigg(1-\frac{12\alpha}{M^2}(k_1^2+k_2^2)\bigg)\times 
\nonumber\\
&&~~~~~\bigg[\Box_4+\partial_{z_1}^2+\partial_{z_2}^2-6A^2(k_1^2+k_2^2)\bigg] 
h_{\mu \nu}  \nonumber\\
&&-\delta(z_1)\bigg[
\frac{8\alpha}{M^2}\frac{k_1}{A}\bigg(\frac{1}{A^2}(\Box_4
+\partial_{z_2}^2) \nonumber \\ 
&&~~~~~+\frac{k_2}{A}sgn(z_2)\partial_{z_2}
-\frac{3k_1}{A}sgn(z_1)\partial_{z_1}\bigg)  \nonumber\\
&&~~~~~+\frac{k_1}{A}\left(10-\frac{\alpha}{M^2}(168k_1^2+152k_2^2)\right)
-\frac{1}{A}\frac{\Lambda_{z_1}}{M^4} 
\bigg]h_{\mu \nu} \nonumber \\
&&-\delta(z_2)\bigg[
\frac{8\alpha}{M^2}\frac{k_2}{A}\bigg(\frac{1}{A^2}(\Box_4
+\partial_{z_1}^2) \nonumber \\
&&~~~~~+\frac{k_1}{A}sgn(z_1)\partial_{z_1}
-\frac{3k_2}{A}sgn(z_2)\partial_{z_2}\bigg)  \nonumber\\
&&~~~~~~+\frac{k_2}{A}\left(10-\frac{\alpha}{M^2}(168k_2^2+152k_1^2)\right) 
-\frac{1}{A}\frac{\Lambda_{z_2}}{M^4}
\bigg]h_{\mu \nu} \nonumber \\
&&-\delta(z_1)\delta(z_2)\bigg[\frac{160\alpha}{M^2}\frac{k_1k_2}{A^2}
-\frac{1}{A^2}\frac{\Lambda_1}{M^4}\bigg]
h_{\mu \nu}=0 ~,  \label{bound0}
\end{eqnarray}
where we use Eq.~(\ref{k1}).   
The above equation for $h_{\mu\nu}$ is more simplified in the conformal 
coordinate,  
\begin{eqnarray}
&-&\frac{1}{2}\bigg(1-\frac{12\alpha}{M^2}k_{\pm}^2\bigg)  
\bigg[\Box_4+\partial_{z_1}^2+\partial_{z_2}^2  \nonumber \\
&&~~~~~-4A\{k_1sgn(z_1)\partial_{z_1}+k_2sgn(z_2)\partial_{z_2}\}\bigg] 
\tilde{h}_{\mu \nu}  \nonumber \\
&&-\frac{\delta(z_1)}{A}\bigg[
\frac{8\alpha}{M^2}k_1\bigg(\Box_4+\partial_{z_2}^2 \nonumber \\
&&~~~~~-3A\{k_1sgn(z_1)\partial_{z_1}+k_2sgn(z_2)\partial_{z_2}\}\bigg)  
\nonumber\\ 
&&~~~~~+A^2\left(8k_1\{1-\frac{12\alpha}{M^2}k_{\pm}^2\}
-\frac{\Lambda_{z_1}}{M^4}\right)
\bigg]\tilde{h}_{\mu \nu} \nonumber \\
&&-\frac{\delta(z_2)}{A}\bigg[
\frac{8\alpha}{M^2}k_2\bigg(\Box_4+\partial_{z_1}^2 \nonumber \\
&&~~~~~-3A\{k_2sgn(z_2)\partial_{z_2}+k_1sgn(z_1)\partial_{z_1}\}\bigg)  
\nonumber \\
&&~~~~~+A^2\left(8k_1\{1-\frac{12\alpha}{M^2}k_{\pm}^2\}
-\frac{\Lambda_{z_2}}{M^4}\right)
\bigg]\tilde{h}_{\mu \nu} \nonumber \\
&&-\delta(z_1)\delta(z_2)\bigg[\frac{96\alpha}{M^2}k_1k_2
-\frac{\Lambda_1}{M^4}\bigg]
\tilde{h}_{\mu \nu}=0 ~,  \label{bound0t}
\end{eqnarray}  
where $\tilde{h}_{\mu\nu}$ is defined in Eq.~(\ref{pcmetric}).  

The bulk contribution in the above equation comes only from 
the first term~(\ref{bound0}). The second and third parts 
of~(\ref{bound0}) and (\ref{bound0t}) 
describe the behavior of the graviton on
the corresponding 4 brane, and the last part of~(\ref{bound0t}) just 
gives a boundary condition of $h_{\mu\nu}$ at the origin(i.e. at the 
3-brane), which is consistent with Eq.~(\ref{k4}).     
In general, the bulk equations, the first part 
of Eq.~(\ref{bound0}) (or (\ref{bound0t})) 
cannot be solved easily, but the solution for the massless mode is 
trivial. If we assume 
$\partial_{z_1}\tilde{h}_{\mu\nu}
=\partial_{z_2}\tilde{h}_{\mu\nu}=0$ and 
put the background relations Eq.~(\ref{k2})--(\ref{k4}) 
into the above equation, we obtain   
\begin{equation}
\Box_4\tilde{h}_{\mu\nu}^0(x)=0~.
\end{equation}
Hence, the massless graviton has the following profile in the bulk, 
\begin{equation}
h^0_{\mu\nu}(x,z_1,z_2)=A^2(z_1,z_2)\tilde{h}^0_{\mu\nu}(x)
=A^2(z_1,z_2)\epsilon_{\mu\nu}e^{ipx}~,
\end{equation} 
where $\epsilon$ is the polarization tensor of the 4 dimensional graviton.  

As the effective 4 dimensional theory would be described  
by the massless graviton dominantly, 
let us calculate the effective 4 dimensional Planck mass 
$M_{P,eff}$ approximately.  
After integrating the extra dimensions with the 4D part of the metric 
as $\tilde{g}_{\mu\nu}(x)\equiv\eta_{\mu\nu}+\tilde{h}_{\mu\nu}$ 
in Eq.~(\ref{metric}), we obtain the 
4D effective action as follows, 
\begin{eqnarray}
S_{eff}&=&\frac{M^2_{P,eff}}{2}\int d^4 x\sqrt{-\tilde{g}^{(4)}}
\bigg[\tilde{R}+\cdot\cdot\cdot\bigg] , \label{effaction1}
\end{eqnarray}
where $\tilde R$ is the 4D Ricci scalar.
The 4D Planck mass is calculated by reading off the coefficients of `$\Box_4$'  
in Eq.~(\ref{bound0}) or (\ref{bound0t}) and integrating those
with respect to $z_1$ and $z_2$, 
\begin{eqnarray}
M^2_{P,eff}&=&M^4\int^\infty_{-\infty} dz_1\int^\infty_{-\infty} dz_2
A^4\bigg[1-\frac{12\alpha}{M^2}k_{\pm}^2 \nonumber \\
&&~~~+\frac{1}{A}\frac{16\alpha}{M^2}
\bigg(k_1\delta(z_1)+k_2\delta(z_2)\bigg)\bigg]\nonumber \\
&=&\frac{2M^4}{3k_1 k_2}\bigg(1+\frac{12\alpha k^2_\pm}{M^2}\bigg) 
\label{pmassein} 
\end{eqnarray}
which gives a finite value. 
Therefore, we can explain gravitational interactions 
consistently even in the non-compact 6 spacetime dimensions.    
Note that our effective 4D Planck mass obtained above
from the Einstein equation is the same
as the one obtained from the action itself by integrating out
$z_1$ and $z_2$, as given in Eq.~(\ref{pmass}).

In case of the absence of the 4-branes, i.e. $k_1^2+k_2^2=M^2/(12\alpha)$, 
the bulk kinetic term in Eq.~(\ref{bound0}) or (\ref{bound0t}) 
does not contribute to the 
linearized Einstein equation and thus the graviton is not allowed to propagate 
in the bulk.
But by higher order terms in the $h_{\mu\nu}$ expansion, 
a certain ``gravity interaction'' could exist in the bulk 
even though the mediating particle cannot be defined as the graviton.  

Now let us discuss the Kaluza-Klein(KK) 
modes of the graviton. We will get a bulk solution first 
using Eq.~(\ref{bound0}) or (\ref{bound0t}), and then apply the 
boundary conditions with the delta functions in the above equations.  
Eq.~(\ref{bound0}) is easier to treat rather than Eq.~(\ref{bound0t}) 
because the former does not have any first derivative terms in the bulk 
equation.
It is possible to separate the variables,  $h_{\mu\nu}(x,z_1,z_2)=
\psi(z_1,z_2)e^{ip\cdot x} \epsilon_{\mu\nu}$, where 
$x^\mu$ and $p^{\mu}$ are 4D coordinate and momentum, 
respectively. Then, the bulk part of Eq.~(\ref{bound0}), which 
is a two dimensional differential equation, is    
$$
\bigg[-\partial_{z_1}^2-\partial_{z_2}^2
+\frac{6(k_1^2+k_2^2)}{(k_1|z_1|+k_2|z_2|+1)^2}\bigg]\psi(z_1,z_2)
$$
\begin{equation}
=m^2\psi(z_1,z_2) ~,\label{kkeq}
\end{equation}
where $p^2=-m^2$.  
To separate the bulk variables, let us introduce a new coordinate ($s,t$),  
\begin{eqnarray}
&s\equiv k_1|z_1|+k_2|z_2|+1\nonumber \\
&t\equiv k_2|z_1|-k_1|z_2|+1.\label{separate}
\end{eqnarray}
Then Eq.~(\ref{kkeq}) becomes 
\begin{equation}
\bigg(k_1^2+k_2^2\bigg)\bigg[-\partial_s^2-\partial_t^2
+\frac{6}{s^2}\bigg]\hat{\psi}(s,t)=m^2\hat{\psi}(s,t) ~,
\end{equation}
where $\hat{\psi}(s,t)\equiv \psi(z_1,z_2)$.  It is separable as  
\begin{eqnarray}
&\bigg[-\partial_s^2+\frac{6}{s^2}\bigg]\phi_s(s)
=m_s^2\phi_s(s) \label{eq1} \\
&-\partial_t^2\phi_t(t)=m_t^2\phi_t(t) ~, \label{eq2}
\end{eqnarray}
where $\phi_s(s)$, $\phi_t(t)$, $m_s^2$ and $m_t^2$ are defined as 
$$
\hat{\psi}(s,t)=\phi_s(s)\phi_t(t)  
$$
\begin{equation}
\frac{m^2}{(k_1^2+k_2^2)}=m_s^2+m_t^2 ~.
\end{equation} From 
Eq.(\ref{eq1}) and (\ref{eq2}), we can see that $m_s^2$, $m_t^2$ and so 
$m^2$ should be positive definite, because they could be regarded as 
a `Hamiltonian' in quantum mechanics, and have 
positive and flat `potentials', respectively. 
Hence, they have positive `energies' or eigenvalues. 
Thus we conclude that {\it there do not exist any tachyonic KK modes}.      

Eqs.~(\ref{eq1}) and (\ref{eq2}) are easily solved and 
have the following solutions, 
\begin{eqnarray}
\phi_s(s)&=&c_1~\sqrt{s}~J_{5/2}(m_ss)+c_2~\sqrt{s}~Y_{5/2}(m_ss) \nonumber \\ 
&=&\sqrt{\frac{2}{\pi m_s}}\bigg[
c_1~\bigg(\bigg(\frac{3}{(m_ss)^2}-1\bigg)~{\rm sin}(m_ss)\nonumber \\
&&-\frac{3}{m_s^2s^2}~{\rm cos}(m_ss)\bigg)  
+c_2~\bigg(\frac{3}{m_ss}~{\rm sin}(m_ss)\nonumber \\
&&+\bigg(\frac{3}{(m_ss)^2}-1\bigg)~{\rm cos}(m_ss)\bigg)\bigg]~, 
\label{bulks}\\
\phi_t(t)&=&d_1~{\rm sin}(m_tt)+d_2~{\rm cos}(m_tt)~,\label{bulkt}
\end{eqnarray}
where $J_{5/2}$ and $Y_{5/2}$ are Bessel functions.   
$c_1$, $c_2$, $d_1$ and $d_2$ are arbitrary constants 
but should be determined by the boundary conditions.  
Note that for large $m_ss$, we have 
\begin{equation}
\phi_s(s)\approx -\sqrt{\frac{2}{\pi}}\bigg[c_1~{\rm sin}(m_ss)
+c_2~{\rm cos}(m_ss)\bigg]~,
\end{equation}
i.e. KK modes behave like free particles.   

On integrating Eq.~(\ref{bound0}) near the extra dimension axes and the origin, 
the boundary conditions for the spin-2 graviton modes are given as follows 
respectively,
$$
\bigg[\bigg(1-\frac{12\alpha k^2_\pm}{M^2}\bigg)\xi
$$
\begin{equation}
+\frac{8\alpha k_1}{M^2 A} (-\xi^\prime
+A(k_2\eta-k_1\xi))\bigg]_{z_1=0+}=0,\label{mbc1}
\end{equation}
$$
\bigg[\bigg(1-\frac{12\alpha k^2_\pm}{M^2}\bigg)\eta
$$
\begin{equation}
+\frac{8\alpha k_2}{M^2 A}
(-\dot{\eta}-A(k_2\eta-k_1\xi))\bigg]_{z_2=0+}=0\label{mbc2},
\end{equation}
\begin{equation}
\frac{8\alpha}{M^2A^3}(k_1\eta+k_2\xi)\mid_{(z_1=0+,z_2=0+)}=0\label{mbc3}
\end{equation}
where we used the bulk equation (\ref{kkeq}) and
\begin{eqnarray}
\xi&=&\psi'+2k_1A\psi \nonumber \\
&=&k_1\bigg(\frac{\partial}{\partial s}+\frac{2}{s}\bigg)\hat{\psi}
+k_2\frac{\partial\hat{\psi}}{\partial t},\label{xi} \\
\eta&=&\dot{\psi}+2k_2A\psi \nonumber \\
&=&k_2\bigg(\frac{\partial}{\partial s}+\frac{2}{s}\bigg)\hat{\psi}
-k_1\frac{\partial\hat{\psi}}{\partial t}\label{eta}.
\end{eqnarray} 
The zero mode solution, $\hat{\psi}_0=A^2=s^{-2}$, is shown to satisfy 
all of the above boundary conditions since $\xi=\eta=0$ identically 
and it is  
regarded as the 4D massless graviton 
since it is a normalizable bound state as its norm being 
$\| \psi_0 \|^2<\infty$. 
For the KK massive modes, 
there are two types of bulk solutions since we have to deal with the zero mode 
separately:
\begin{eqnarray}
\hat{\psi}^{(1)}_m&=&s^{-2}\phi_t(t) \nonumber \\
&=&s^{-2}(d_1\sin(m_t t)+d_2 \cos(m_t t)),\label{type1} \\
\hat{\psi}^{(2)}_m&=&\phi_s(s)\phi_t(t) \nonumber \\ 
&=&\sqrt{s}(c_1 J_{5/2}(m_s s)+c_2 Y_{5/2}(m_s s)) \nonumber \\
&&\cdot(d_1\sin(m_t t)+d_2 \cos(m_t t))\label{type2}
\end{eqnarray}
where $\phi_s$ and $\phi_t$ are given by Eqs.~(\ref{bulks}) and (\ref{bulkt}), 
respectively and we note that $m^2_s=0$ for the case of $\hat{\psi}^{(1)}_m$.

Then, $\hat{\psi}^{(1)}_m$ satisfies the boundary condition 
at the origin automatically for $k_1=k_2$ but 
otherwise only with $d_2/d_1=\cot(m_t)$. 
And the remaining boundary conditions are rewitten as
\begin{equation}
\bigg[\bigg(1-\frac{12\alpha k^2_\pm}{M^2}\bigg)\frac{d\phi_t}{dt}+\frac{8\alpha
k_1 k_2 m^2_t}{M^2} s\phi_t\bigg]\mid_{z_1=0+}=0,\label{t1b1}
\end{equation}
\begin{equation}
\bigg[\bigg(1-\frac{12\alpha k^2_\pm}{M^2}\bigg)\frac{d\phi_t}{dt}-\frac{8\alpha
k_1 k_2 m^2_t}{M^2} s\phi_t\bigg]\mid_{z_2=0+}=0.\label{t1b2}
\end{equation} 
There exist no KK massive modes of type $\hat{\psi}^{(1)}_m$ satisfying 
the above boundary conditions. 
On the other hand, for the KK massive modes of the other type 
$\hat{\psi}^{(2)}_m$, the boundary 
conditions look complicated to solve, but with assuming no $t$ dependence, 
we can obtain the ratio between coefficients of the Bessel functions as 
\begin{eqnarray}
\frac{c_1}{c_2}=-\frac{Y_{3/2}(m_s)}{J_{3/2}(m_s)}\label{ratio},
\end{eqnarray} 
and the boundary conditions on the extra dimension axes are simplified as 
$$
\bigg[\bigg(1-\frac{\alpha}{M^2}(36k_1^2+4k^2_2)\bigg)\bigg(\frac{d}{ds}
+\frac{2}{s}\bigg)
\phi_s
$$
\begin{equation}
+\frac{8\alpha k_1^2 m_s^2}{M^2}s\phi_s\bigg]|_{z_1=0+}=0, \label{t2b1}
\end{equation}
$$
\bigg[\bigg(1-\frac{\alpha}{M^2}(4k_1^2+36k^2_2)\bigg)\bigg(\frac{d}{ds}
+\frac{2}{s}\bigg)
\phi_s
$$
\begin{equation}
+\frac{8\alpha k_2^2 m_s^2}{M^2}s\phi_s\bigg]|_{z_2=0+}=0.\label{t2b2}
\end{equation} 
However, the above boundary conditions are not satisfied by the KK massive 
modes of a function of $s$ only except for $m^2_s=0$, i.e., the zero mode. 
Moreover, the situation would 
not be different for the more general KK modes of type $\hat{\psi}^{(2)}_m$. 
Therefore, even though the bulk equation for the 4D massive gravitons 
is exactly solvable, 
there would not exist bulk solutions satisfying the boundary conditions along 
the extra dimension axes with the simple ansatz for separation of variables, 
Eq.~(\ref{separate}). It is shown that this situation does not become different 
even without the Gauss-Bonnet term.

\section{confining gravity to the solitonic 4-branes}

Let us discuss the case with the orthogonal 4-branes being regarded as 
{\it solitonic} by choosing the relation 
between bulk parameters Eq.~(\ref{bulktune}), 
for which there is no six-dimensional bulk propagation of 
graviton but the gravity is confined to the solitonic 4-branes as shown 
in Eq.~(\ref{bound0}) or (\ref{bound0t}). In this case, we can rewrite the 
linearized equation (\ref{bound0t}) with $\tilde{h}_{\mu\nu}
=A^{-3/2}\tilde{\psi}(z_1,z_2) e^{ip\cdot x}\epsilon_{\mu\nu}$ as
$$
-\delta(z_1)\frac{8\alpha k_1}{M^2}\bigg[m^2+\partial^2_{z_2}
-\frac{15}{4}k^2_2 A^2+3k_2 A\delta(z_2)\bigg]\tilde{\psi} 
$$
$$
+\frac{24\alpha k_1^2}{M^2} sgn(z_1)A\delta(z_1)(\partial_{z_1}
+\frac{3}{2}k_1 A)\tilde{\psi}
$$
$$
-\delta(z_2)\frac{8\alpha k_2}{M^2}\bigg[m^2+\partial^2_{z_1}
-\frac{15}{4}k^2_1 A^2+3k_1 A\delta(z_1)\bigg]\tilde{\psi}
$$
\begin{equation}
+\frac{24\alpha k_2^2}{M^2} sgn(z_2)A\delta(z_2)(\partial_{z_2}
+\frac{3}{2}k_2 A)\tilde{\psi}=0.\label{confine}
\end{equation}
 
Then, the above equation is decomposed into two five-dimensional bulk 
equations of graviton and three boundary conditions:
\begin{eqnarray}
\bigg(-\partial^2_{z_1}+\frac{15}{4}k^2_1 A^2\bigg)\tilde{\psi}
&=&m^2\tilde{\psi},({\rm along \, {\it z_1}\, axis}) \label{f1}\\
\bigg(-\partial^2_{z_2}+\frac{15}{4}k^2_2 A^2\bigg)\tilde{\psi}
&=&m^2\tilde{\psi}, ({\rm along \, {\it z_2}\, axis})\label{f2} 
\end{eqnarray}
\begin{eqnarray}
\bigg(\partial_{z_{1}}+\frac{3}{2}k_{1}A\bigg)\tilde{\psi}|_{z_{1}=0+}&=&0, 
\label{fb1}\\
\bigg(\partial_{z_{2}}+\frac{3}{2}k_{2}A\bigg)\tilde{\psi}|_{z_{2}=0+}&=&0, 
\label{fb2}
\end{eqnarray}
\begin{equation}
\bigg[\bigg(\partial_{z_{1}}+\frac{3}{2}k_{1}A\bigg)\tilde{\psi}
+\bigg(\partial_{z_{2}}+\frac{3}{2}k_{2}A\bigg)
\tilde{\psi}\bigg]|_{(z_1=z_2=0+)}=0\label{fb3}
\end{equation}
where we note that the last equation is a necessary consequence 
in case that the third and fourth ones are satisfied and vice versa 
for our case as will be shown later. From 
Eqs.~(\ref{f1}) and (\ref{f2}), the zero mode 
solution for $m^2=0$ becomes the same as in the non-solitonic case,
\begin{equation}
\tilde\psi_0=(k_1|z_1|+k_2|z_2|+1)^{-3/2}
\end{equation}
which automatically satisfies the boundary conditions, 
Eqs.~(\ref{fb1}-\ref{fb3}). Note that the zero mode wave $\tilde{\psi}_0$ 
is chosen to be nonvanishing only along the solitonic 4-branes.

On the other hand, solving Eqs.~(\ref{f1}) and (\ref{f2}), 
the KK mode solutions 
are given as linear combinations
of Bessel functions of order two as in the RS case, propagating along solitonic 
4-branes located at the $z_1$ and $z_2$ axes:
\begin{eqnarray}
\tilde{\psi}_m&=&N^{(1)}_m(|z_1|+1/k_1)^{1/2}
[Y_2(m(|z_1|+1/k_1)) \nonumber \\
&&+B_m J_2(m(|z_1|+1/k_1))],({\rm along \, {\it z_1} \, axis})
\end{eqnarray}
\begin{eqnarray} 
\tilde{\psi}_m&=&N^{(2)}_m(|z_2|+1/k_2)^{1/2}
[Y_2(m(|z_2|+1/k_2))\nonumber \\
&&+C_m J_2(m(|z_2|+1/k_2))],({\rm along \, {\it z_2} \, axis})
\end{eqnarray}
where $N^{(1,2)}_m$, $B_m$ and $C_m$ are constants to be determined by boundary 
conditions and normalization. Then, for the KK modes with small masses, 
i.e., $m(|z_{1,2}|+1/k_{1,2})\ll 1$, the constants $B_m$ and $C_m$ are 
determined approximately 
from the boundary conditions, Eqs.~(\ref{fb1}) and (\ref{fb2}), 
as the following,
\begin{equation}
B_m\simeq\frac{4k^2_1}{\pi m^2}, \ \ \ C_m\simeq\frac{4k^2_2}{\pi m^2}.
\end{equation}
Furthermore, from the plane wave normalization such that
\begin{eqnarray}
1=\int_0^{z_c}dz_{1}|\tilde{\psi}_m|^2+\int_0^{z_c}dz_{2}|\tilde{\psi}_m|^2,
\end{eqnarray}
we also obtain the normalization constant $N^{(1,2)}_m$ as
\begin{equation}
N^{(1)}_m\sim B_m^{-1}\sqrt{\frac{\pi m}{z_c}}\bigg(1+\frac{k_2}{k_1}\bigg)^{-1/2}
=\bigg(\frac{k_2}{k_1}\bigg)^{3/2}N^{(2)}_m.
\end{equation}
Therefore, the Newtonian potential for two point sources $m_1,m_2$ 
separated by $r$ on the 3-brane is found in a conventional way to be 
\begin{eqnarray}
V(r)&\simeq& \frac{G_N m_1 m_2}{r} \nonumber \\
&+&(16\alpha k_2 M^2)^{-1}\int_0^\infty dm 
\frac{m_1 m_2 e^{-mr}}{r}|\tilde{\psi}_m(0)|^2 \nonumber \\
&+&(16\alpha k_1 M^2)^{-1}\int_0^\infty dm 
\frac{m_1 m_2 e^{-mr}}{r}|\tilde{\psi}_m(0)|^2 \nonumber \\
&\simeq&\frac{G_N m_1 m_2}{r}\bigg[1+\bigg(\frac{k^2_\pm}{k_1 k_2}\bigg)^2
\frac{1}{(k_\pm r)^2}\bigg]
\end{eqnarray}
where we used $G_N=M^{-2}_P=(3k_1 k_2)/(4M^4)$ from Eq.~(\ref{pmass}), 
$|\tilde{\psi}_m(0)|^2\sim m/(k_1+k_2)$ and the effective 5D gravity 
couplings for KK modes are read off from coefficients of the 5D kinetic terms 
in Eq.~(\ref{confine}).
As a result, 
corrections due to the KK massive modes are {\it five-dimensional} due to the 
confinement of gravity to the solitonic 4-branes and 
suppressed in comparison with the 
Newton force at larger length scales than the curvature scales.
Consequently, the confinement of gravity exactly gives rise to two copies of 
the five-dimensional RSII model. In addition, since gravity does not propagate 
into the bulk, one fine-tuning condition between bulk parameters, 
Eq.(\ref{bulktune}), remains intact at the quantum level of linearized gravity.

\section{The mass hierarchy with the orbifold $T^2/(Z_2\times Z_2)$ }

We have just shown that there exists two orthogonal 4-brane solution with the 
nonzero tension of the intersection (or 3-brane) in 6D with  
the Gauss-Bonnet term.
Therefore, it is possible to put another 3-brane in the appropriate position 
of the bulk as the additional intersection of 4-branes to solve the hierarchy 
problem as in the RS I case. But, it should 
be guaranteed that the additional brane should be located at the fixed point
of the orbifold to be stable, i.e., the bulk should end at the position of the 
additional brane. Thus, we assume that there exist the compact extra 
dimensions with the orbifold $T^2/(Z_2\times Z_2)$, where $Z_2$ acts 
on each extra dimension once. And let us set the range of the extra coordinates
as $z_1\in (-a,a)$ and $z_2\in (-b,b)$. Here we assumed the periodicity 
of $2a(2b)$ along $z_1(z_2)$ direction.
Then, with the $Z_2\times Z_2$ symmetric solution Eq.~(\ref{sol}), 
we need four 3-branes to match the boundary conditions at the four fixed 
points of the torus, $(z_1,z_2)=(0,0)$, $(a,0)$, $(a,b)$ and $(0,b)$. 
Let us denote the 3-brane tensions as $\Lambda_1$, $\Lambda_2$, $\Lambda_3$ 
and $\Lambda_4$ in order. And the neighboring two 3-branes are connected 
to each other by one 4-brane denoted as $\Lambda_{12}$, $\Lambda_{23}$, 
$\Lambda_{34}$ and $\Lambda_{41}$ in cyclic order.  
If the boundary equations in Eqs.~(\ref{zero})-(\ref{six}) are changed into 
the following, 
\begin{equation}
e_1=-M^{-4}\frac{1}{A}(\Lambda_{41}\delta(z_1)
+\Lambda_{23}\delta(z_1-a)),\label{z2dir}
\end{equation}
\begin{equation}
e_2=-M^{-4}\frac{1}{A}(\Lambda_{12}\delta(z_2)
+\Lambda_{34}\delta(z_2-b)),\label{z1dir}
\end{equation}
\begin{equation}
e_3=-M^{-4}\sum_{i=1}^4\frac{1}{A^2}\Lambda_i\delta(z_1-z^{(i)}_1)
\delta(z_2-z^{(i)}_2),
\label{sum}
\end{equation}
where $z^{(i)}_1,z^{(i)}_2$ are positions of the branes,
then we obtain the following relations between the 4-brane tensions
and similarly for the 3-brane tensions,
\begin{equation}
\Lambda_{41}=-\Lambda_{23}=k_1\bigg(1-\frac{12\alpha k^2_\pm}{M^2}\bigg),
\end{equation}
\begin{equation}
\Lambda_{12}=-\Lambda_{34}=k_2\bigg(1-\frac{12\alpha k^2_\pm}{M^2}\bigg),
\end{equation}
\begin{equation}
\Lambda_1=\Lambda_3=-\Lambda_2=-\Lambda_4=96\alpha k_1 k_2 M^2. 
\label{tens}
\end{equation}  
In general, in view of Eqs.~(\ref{k1}-\ref{k4}), for fixed bulk parameters, 
two orthogonal 4-brane tensions should be fine-tuned with the 3-brane tension 
on their intersection (e.g., between $\Lambda_{41}(\Lambda_{12})$ and 
$\Lambda_1$ and etc.).    
When we adopt the string solution with two {\it solitonic} 4-branes, 
each 3-brane tension can 
take an arbitrary value of either sign irrespective of the bulk 
parameters as argued in the previous section, but it should be fine-tuned 
to one another as shown in Eq. (\ref{tens}). 
Then, to explain the 
large mass hierarchy for both the string solution with {\it non-solitonic} 
4-branes for 
$\alpha>0$ and the string solution with {\it solitonic} 4-branes, 
we may take the $\Lambda_3$ brane with positive tension as the visible brane, 
whereas the $\Lambda_1$ brane can be considered as the hidden brane
of the Planck scale. 
In addition, if $\Lambda_2$ brane and 
$\Lambda_4$ branes are considered as the
second and the third generation family branes while the $\Lambda_3$
brane is interpreted as the first family brane, we may understand
the mass hierarchy between families and neutrino oscillation. 
In this case, the gauge fields are required to live in the bulk.
But, we do not digress into this family problem here. 

Before considering how the mass hierarchy is generated in this model, 
let us rewrite the metric as
\begin{eqnarray}
ds^2_6&=&A^2(z_1,z_2)(\eta_{\mu\nu}dx^\mu dx^\nu+dz_1^2+dz_2^2)\nonumber\\
&=&A^2(y_1,y_2)\eta_{\mu\nu}dx^\mu dx^\nu+B^2(y_1,y_2)dy_1^2\nonumber\\
&+&C^2(y_1,y_2)dy_2^2
\end{eqnarray}
by the following bulk coordinate transformations:
\begin{eqnarray}
dz_1=\frac{B}{A}dy_1,\,\,dz_2=\frac{C}{A}dy_2,
\end{eqnarray}
i.e. $k_1z_1=sign(y_1)(e^{k_1|y_1|}-1),k_2z_2=sign(y_2)
(e^{k_2|y_2|}-1)$.
Then, we can have the metric functions in the new coordinate: 
$A=(e^{k_1|y_1|}+e^{k_2|y_2|}-1)^{-1}$, $B=e^{k_1|y_1|}A$ 
and $C=e^{k_2|y_2|}A$. 
So, the 4D Planck mass becomes 
\begin{eqnarray}
M^2_{P,eff}&=&M^4\int^a_{-a} dz_1\int^b_{-b} dz_2
\bigg[A^4\bigg(1+\frac{12\alpha}{M^2}\frac{1}{A^2}
\bigg(\bigg(\frac{A'}{A}\bigg)^2\nonumber \\
&+&\bigg(\frac{\dot{A}}{A}\bigg)^2\bigg)\bigg) 
-\frac{12\alpha}{M^2}\bigg((AA')'+(A\dot{A})^{\dot{\ }}\bigg)\bigg]
\nonumber\\
&=&M^4\bigg(1+\frac{12\alpha k^2_\pm}{M^2}\bigg)
\int_{-b_1}^{b_1}dy_1\int_{-b_2}^{b_2}dy_2 A^2BC \nonumber\\
&=&\frac{2M^4}{3k_1 k_2}\bigg(1+\frac{12\alpha k^2_\pm}{M^2}\bigg)
\bigg[1
+(e^{k_1b_1}+e^{k_2b_2}-1)^{-2}\nonumber \\
&-& e^{-2k_1b_1}-e^{-2k_2b_2}
\bigg]
\end{eqnarray}
where $(AA')'$ and $(A\dot{A})^{\dot{\ }}$ terms in the first line vanish 
after integration due to the periodicity of the extra dimensions and 
$b_1,b_2$ are the range of the extra dimensions in the new coordinate 
and in the limit of $b_1\rightarrow \infty$ and $ b_2\rightarrow \infty$, 
Eq.~({\ref{pmass}) can be reproduced. Note that the 4D Planck mass has 
a finite value if $k_1k_2\neq 0$,
i.e., $\Lambda_i \neq 0$ for all $i$ from Eqs.~(\ref{k4}) and (\ref{sum}) 
and its positiveness is assured for $|\alpha|<\frac{M^2}{12k^2_\pm}$ for 
$\alpha<0$ and any value for $\alpha>0$.
In this new coordinate, let us consider the action for the Higgs scalar field 
at the $\Lambda_3$ brane,
\begin{eqnarray}
&S_{vis}\supset\int dx^4 \sqrt{-g^{(vis)}}\bigg[\bar{g}^{\mu\nu}
\partial_\mu H\partial_\nu H-(H^2-m_0^2)^2\bigg],\nonumber \\
&=\int dx^4 \sqrt{-g^{(4)}}A^4\bigg[A^{-2}(\partial H)^2-(H^2-m_0^2)^2\bigg],
\end{eqnarray}  
which becomes of a canonical form by redefining 
the scalar field as $\tilde{H}=AH$, 
\begin{eqnarray}
\int dx^4\sqrt{-g^{(4)}}\bigg[(\partial \tilde{H})^2
-(\tilde{H}^2-m^2_3)^2\bigg]
\end{eqnarray}
where the Higgs mass parameter on the visible brane is given by
\begin{eqnarray}
m_3=Am_0=(e^{k_1 b_1}+e^{k_2 b_2}-1)^{-1}m_0.
\end{eqnarray}
Similarly, we obtain the effective mass scales on the other branes, 
$\Lambda_2$ and $\Lambda_4$, respectively:
\begin{eqnarray}
m_2=e^{-k_1 b_1}m_0,\ \ \ m_4=e^{-k_2 b_2}m_0.
\end{eqnarray}
Therefore, when we regard the $\Lambda_3$ brane as our universe, 
we can obtain the hierarchy between the Planck scale($m_0$) 
and the weak scale($m_3$) by choosing $k_1 b_1$ and/or $k_2 b_2$ as about 37. 
It is interesting to see that the mass parameters on the branes are related by
\begin{eqnarray}
\frac{1}{m_2}+\frac{1}{m_4}-\frac{1}{m_3}=\frac{1}{m_0}.\label{massrel}
\end{eqnarray}
where $m_0$ is the mass scale of order the Planck mass at the
3-brane located at $(0,0)$. Since
the RHS of Eq. (\ref{massrel}) is negligible, the magnitudes of
at least two of $m_2, m_3$ and $m_4$ are of the same order, 
which may allow to a deeper understanding of the family structure.
Instead of putting different families in the different 3-branes, one
can put all the fermions and the Higgs doublet in the $(a,b)$
brane or in the $(a,b)$ and $(0,b)$ branes with $b\gg a$.
Then the $(a,0)$ brane can be used for an intermediate scale
brane. However, it is not necessarily needed as proposed in \cite{kyae} 
toward a solution of the $\mu$ problem with supersymmetry\cite{mu},
because the visible sector fields here are 
already put at the TeV brane. 
On the other hand, if the visible sector fields with supersymmetric
extension are put at the
two Planck scale branes at $(0,0)$ and $(a,0)$ with $b\gg a$,
then it is needed to introduce intermediate scale brane(s) at
$(0,b)$ and $(a,b)$\cite{kyae}. In this case, there can be two 
intermediate scales in principle due to the two three 
branes at the intermediate scales.

\section{conclusion}

In this paper we obtained the localized gravity on the intersection 
of two orthogonal {\it non-solitonic} or {\it solitonic} 4-branes in the
Einstein-Gauss-Bonnet theory in 6D.
The nonzero 3-brane tension is allowed, which has been possible 
due to the presence of the Gauss-Bonnet term.
The Gauss-Bonnet term can contain a product of
two terms with two derivatives of the metric on each 
term. Therefore, in the EGB theory 3-brane solutions 
are not possible beyond 6D. To have 3-brane solutions
beyond 6D, we have to introduce higher derivative gravity 
than the Gauss-Bonnet term. 

The solution has a warp factor which decreases exponentially
at large distance from the origin in the extra dimension.
If the $Z_2\times Z_2$ symmetry is assumed on the bulk space even without 
{\it non-solitonic} 4-branes, 
one can consider a solution of a 3-brane residing on the intersection of 
two {\it solitonic} 4-branes
for the localization of gravity and also for
a possible solution of the cosmological constant problem
as in the RSII model\cite{KKL2}. With this solution, it is interesting to 
make the confinement of
gravity to the solitonic 4-branes possible, which results in nothing but 
two copies of the 5D RSII model. 
In addition, the extra dimension can be compactified.
The $T^2/(Z_2\times Z_2)$ orbifold symmetry gives four
fixed points where 3-branes resides on intersections of two 4-branes.
In this case, the electroweak scale versus the Planck scale
hierarchy can be understood. We also pointed out  
the possibility of understanding the family structure, which
will be studied in a future publication.

\acknowledgments
This work is supported in part by the BK21 program of Ministry 
of Education, CTP Research Fund of Seoul National University,
and by the Center for High Energy Physics(CHEP),
Kyungpook National University.

\end{document}